\documentclass[submission,copyright,creativecommons]{eptcs}
\providecommand{\event}{ACL2 2014}
\usepackage{breakurl}             
\usepackage{alltt, graphicx,latexsym,pifont,amsmath,amstext,amssymb,amsfonts,epic,eepic}

\title{Modeling Algorithms in SystemC and ACL2}
\author{John W.~O'Leary
\institute{Intel Corporation\\Formal Verification Center of Expertise\\Hillsboro, Oregon, USA}
\email{john.w.oleary@intel.com}
\and
David M.~Russinoff
\institute{Intel Corporation\\Atom Products Division\\Austin, Texas, USA}
\email{david.m.russinoff@intel.com}
}
\def\titlerunning{MASC and ACL2}
\def\authorrunning{J.~W.~O'Leary \& D.~M.~Russinoff}
\begin{document}
\maketitle

\begin{abstract}
We describe the formal language MASC, based on a subset of SystemC and intended for modeling algorithms to be implemented in hardware.  
By means of a special-purpose parser, an algorithm coded in SystemC is converted to a MASC model
for the purpose of documentation, which in turn is translated to ACL2 for formal verification.  The parser also generates a SystemC
variant that is suitable as input to a high-level synthesis tool.  As an illustration of this methodology, we
describe a proof of correctness of a simple 32-bit radix-4 multiplier.
\end{abstract}

\section{Introduction}\label{intro}

Formal verification of hardware designs is typically applied to register-transfer logic (RTL) models coded in Verilog,
and therefore cannot begin until stable RTL is available.  Since bugs are often introduced at the algorithmic
level much earlier in the design process, a hierarchical approach to design verification is desirable, beginning with a
comprehensive mathematical proof of correctness of the underlying algorithm, perhaps conducted in parallel 
with RTL development, and completed by demonstrating that the algorithm is faithfully implemented in RTL.  
However, the details of an arithmetic algorithm may be difficult to ascertain.  The description
provided by an architect is often inscrutable to verifiers and designers alike.  Even when an executable model is provided,
it is not readily susceptible to formal analysis, nor is there any straightforward way of comparing it 
with the derived RTL.

At Intel, we have proposed the adoption of a limited subset of SystemC~\cite{systemc}, an ANSI standard C++ class library
intended for system and hardware design, as the basis of a standard modeling language
to be shared by architects, designers, and verification engineers in the specification, implementation,
and formal verification of arithmetic algorithms.  The choice of language, the particular
restrictions that we impose on it, and the extensions that we provide are driven by the following objectives:

\begin{itemize}

\item Documentation: C++ is a natural candidate in view of its versatility and widespread use 
in system modeling.  For our purpose as a specification language, we require a subset that is
simple enough to allow a clear and easily understood semantic definition, but sufficiently expressive
for detailed encoding of complex arithmetic algorithms.

\item RTL development:  This is the motivation for incorporating the SystemC library, which includes
a set of data types that model integer and fixed-point registers of arbitrary width and provide the basic 
bit manipulation features of Verilog, thereby closing the semantic gap between an algorithm and its RTL 
implementation.

\item Formal analysis: The goal of susceptibility to direct mathematical analysis as well as formal 
reasoning tools dictates a functional programming style, which we promote by eliminating side-effects,
replacing the pointers and reference parameters of C++ with other suitable extensions. The objective
is a language that is susceptible to translation to ACL2.

\end{itemize}

Detailed modeling of hardware {\em implementations} is not among our
objectives; our SystemC subset, therefore, does not include constructs that describe
combinatorial gates, sequential elements, or their
interconnection. Instead, derivation of hardware implementations is achieved
with high-level synthesis tools, such as Cadence Design System's
C-to-Silicon compiler (CtoS)~\cite{ctos}, which convert SystemC to
Verilog. As a proof of concept we have applied CtoS to several complex
algorithms modeled in our subset and obtained RTL that is
clock-cycle-accurate with respect to to hand-written Verilog code implementing the
same algorithms.

A major selling point of our approach is that arithmetic architects can describe their algorithms in a proper
subset of SystemC and simulate them using standard SystemC environments, a
variety of which are available as open source or from commercial
vendors. As a bridge to formal analysis we have defined an abstract variant of this subset called {\it MASC} (Modeling Algorithms in SystemC), 
with a simpler syntax and semantics.  For example, the format of the MASC bit manipulation operators is similar to 
that of Verilog and more readable than the syntax of C++ methods.  An important semantic dictinction is that MASC 
arithmetic, like ACL2, is unbounded and arbitrarily precise, whereas in SystemC this is true only for some of the register 
classes.  However, the two languages are closely related and have been designed in parallel to ensure that (a)~every 
program in the SystemC subset is readily translatable to MASC, (b)~every MASC program can be so generated from a 
SystemC program, and (c)~every MASC program can be translated to ACL2.

Thus, three distinct representations of an algorithm are provided to serve a variety of purposes: (1)~the hand-coded
SystemC model may be directly executed or synthesized by an existing tool; (2)~the derived MASC version may serve
as documentation or as a reference model for mathematical analysis; and (3)~the ACL2 translation enables formal 
verification. We have implemented a special-purpose parser for our SystemC subset, which performs the following functions:

\begin{itemize}

\item Following a check to ensure that a model conforms to the prescribed restrictions, a MASC translation is generated.

\item Since some of the requirements of high-level synthesis are in conflict with those of readability and formal analysis,
a model that is intended for synthesis is transformed to another variant of C++ that
is suitable as input to CtoS.

\item An S-expression representation of the model is generated.  This is a first step toward translation to ACL2,
which is completed in ACL2 itself. 

\end{itemize}
In this paper, we focus on the implications of the last of these features and the process of proving correctness of
MASC models.  

The problem of translating imperative programs to recursive functions for the purpose of formal verification was first explored 
by McCarthy~\cite{mccarthy} and has been the subject of a number of recent investigations~\cite{peter,myreen}.  None of 
these, however, has been successfully applied to a language suitable for modeling arithmetic circuits as complex as the applications
that we have addressed, which include an aggressively high-performance division and square root algorithm developed in
collaboration with an arithmetic architecture group.

Our proof methodology is based on a library of ACL2 books developed at Advanced Micro Devices, Inc. over 
the course of fifteen years in support of the formal verification of the floating-point units of AMD's
line of microprocessors~\cite{imacs}.  It contains over 600 lemmas pertaining to bit vectors and logical operations,
floating-point representations and arithmetic, and special-purpose techniques relevant to the implementation of elementary
arithmetic operations.  The latest version of the library includes a book that provides a connection
with Centaur Technology's GL tool~\cite{gl}, which allows intermediate results that are susceptible to symbolic
simulation to be proved automatically.  The library belongs to the standard ACL2 release, residing in the Community Books 
directory \verb!"books/rtl/rel9/lib/"!, and is documented in an on-line reference manual~\cite{libman}.

In the following sections, we describe the MASC language, its relation to SystemC, and the process of 
transation to ACL2.  As an illustration, we present a MASC model of a basic radix-4 Booth multiplier, designed for the purpose 
of illustrating our methodology.  This design is intended to be simple enough to be readily comprehended but sufficiently rich
to demonstrate the application of the RTL library in conjunction with GL, and to illustrate some of the problems encountered
in the analysis of ACL2 functions derived from iterative programs.
A set of related files may be found in the ACL2 books directory under \verb!"workshops/2014/russinoff-oleary/support/"!.\footnote{These 
supporting materials are compatible with released versions of ACL2 strictly after Version 6.4, and with svn snapshots of ACL2 and the 
Community Books starting with acl2-devel svn revision 1301 and acl2-books svn revision 2763, respectively.}

\section{MASC: The Formal Language}\label{masc}

Our formal modeling language is designed to correspond as closely as possible to a subset of SystemC.  However,
we draw a fundamental distinction between the two languages, which are described separately in this section and the next.

\subsection{Program Structure}

A MASC program consists of function definitions, type declarations,
and global constant declarations (global variables are not permitted), all of which have the same
syntax as C with the exceptions noted below.

The statements that compose the body of a function may include local type declarations, local variable
and constant declarations, assignments, assertions (which have no semantics), and control 
statements corresponding to the keywords {\tt if}, {\tt for}, {\tt while}, and {\tt switch}.

\subsection{Data Types}

The MASC data types include the primitive numerical types {\tt bool} (boolean values), {\tt uint} (unsigned 
integers), and {\tt int} (signed integers) of C, as well as arrays, structures ({\tt struct}), and enumerations
({\tt enum}).  In a departure from standard C, {\tt uint} and {\tt int} values are in principle unbounded, but in practice,
these types are used only for small (64-bit) numbers.  

Also included are a set of integer and fixed-point register types that roughly correspond to the class templates of
SystemC.  For every positive integer $n$, {\tt ui}$n$ and {\tt si}$n$
are the unsigned and signed integer register types of width $n$, respectively; for $n \geq m > 0$, {\tt uf}$n${\tt i}$m$
and {\tt sf}$n${\tt i}$m$ are the unsigned and signed fixed-point types of width $n$ with $m$ integer bits and $n-m$ 
fractional bits.  Two values are associated with a register: the {\it raw value} of a register of any type of width $n$ is a 
bit vector of width $n$, which in our theory is an integer in the interval $[0, 2^n)$; its {\it interpreted value} is an 
integer or rational number derived from the raw value in the expected manner according to the variable's type.

\subsection{Primitive Operations}

Most of the basic numerical operators of C are provided, with the notable exception of division.
All arithmetic operations are performed in unbounded and arbitrary-precision rational arithmetic.
When the value $v$ of an expression is assigned to a variable, it is modified according the variable's type:

\begin{itemize}

\item {\tt uint} or {\tt int}: The fractional part of $v$ is discarded.  The resulting
value of the variable is $\lfloor v \rfloor$.

\item {\tt ui}$n$ or {\tt si}$n$: The fractional part of $v$ is discarded along with the most
significant integer bits as dictated by the destination format.  The resulting
raw value of the variable is $\lfloor v \rfloor \bmod 2^n$.

\item {\tt uf}$n${\tt i}$m$ or {\tt sf}$n${\tt i}$m$: the least significant fractional bits and the most
significant integer bits of $v$ are discarded as dictated by the destination format.  The resulting
raw value of the variable is $\lfloor 2^{n-m}v  \rfloor \bmod 2^n$.

\end{itemize}
The bitwise logical operators of C are also provided, but may be applied only to registers.  The same is true of
the bit {\it subrange} operator, $x[i:j]$, where x is a (fixed-point or integer) register of width $n$ and $i$ 
and $j$ are integers with $0 \leq j \leq i < n$, as well as the {\it bit reference} operator, $x[i] = x[i:i]$.
These operators are applied only to the raw value of a register, without regard for its type.
They may occur either within an expression to be evaluated or as the object of an assignment.

\subsection{Function Parameters}\label{functions}

All function arguments, including arrays, are passed by value.  Thus, MASC functions are free of side-effects.
In order to compensate for the absence of pointers and reference parameters, a function may return
several values of arbitrary types, in which case the return type is replaced by a list of types,
{\tt <$T_1,\ldots,T_k$>}.  A value returned by such a function is similarly specified as a list of
expressions.  While the same effect could be achieved by
an ordinary {\tt struct} return type, this feature provides a convenient means of
simultaneously assigning the components of a returned value to local variables of the caller.

For example, the following function performs integer division and returns a quotient and remainder as a multiple value:
\begin{verbatim}
  <uint, uint> Divide(m uint, n uint) {
    assert(n != 0);
    uint quot = 0, rem = m;
    while (rem >= n) {
      quot++;
      rem -= n;
    }
    return <quot, rem>;
  }
\end{verbatim}
The result of a division may be recorded as follows:
\begin{verbatim}
  uint q, r;
  <q, r> = Divide(23, 5);
\end{verbatim}

\subsection{Control Restrictions}\label{control}

A number of restrictions are imposed on the syntax of MASC control statements in order to facilitate
the transformation from an imperative to a functional paradigm.  In particular, the ACL2 translator converts a 
{\tt for} loop into an auxiliary recursive function.  In order to ensure the admissibility of this function,
a loop that is to be translated to ACL2 has the form

\begin{alltt}
  for ({\it init}; {\it test}; {\it update}) \{ ... \}
\end{alltt}
where

\begin{itemize}

\item {\it init} is an initialization of a single integer ({\tt int} or {\tt uint}) loop variable.

\item {\it test} is either a comparison between the loop variable and a numerical expression
of the form {\it var} {\it op} {\it limit}, where {\it op} is \verb!<!, \verb!<=!, \verb!>!, 
or \verb!>=!, or a conjunction of the form $\mbox{\it test}_1$ \verb!&&! $\mbox{\it test}_2$, where $\mbox{\it test}_1$ 
is such a comparison (and $\mbox{\it test}_2$ may be any numerical expression) .

\item {\it update} is an assignment to the loop variable.  The combination of {\it test} and {\it update} must
guarantee termination of the loop.  The translator derives a {\tt :measure} declaration from {\it test}, which is
used to establish the admissibility of the generated recursive function.

\end{itemize}
Neither {\tt break} nor {\tt continue} may occur in a {\tt for} loop. In some  cases, the loop
test may be used to achieve the functionality of {\tt break}. For example, instead of

\begin{alltt}
  for (uint i=0; i<N; i++) \{if ({\it expr}) break; ... \}
\end{alltt}
we may write
\begin{alltt}
  for (uint i=0; i<N && \verb@!@{\it expr}; i++) \{ ... \}
\end{alltt}

Although {\tt while} statements cannot be handled directly by the translator, we shall describe in Subsection~\ref{systemc}
a mechanism for accommodating {\tt while} as well as non-compliant {\tt for} loops when a bound on the number of
iterations can be asserted.

Further restrictions are imposed on the placement of {\tt return}
statements.  We require every MASC function body to satisfy the following definition:
A statement block is {\it well-formed} with respect to {\tt return} statements if (1) it consists of 
a non-empty sequence of statements, (2) none of these statements except the final one contains a 
{\tt return} statement, and (3) the final statement of the block is either a {\tt return} statement or an 
{\tt if}\dots{\tt else} statement of which both branches are well-formed with respect to {\tt return} 
statements.

\section{Mapping SystemC to MASC}\label{systemc}

In this section, we discuss the syntactic differences that must be addressed in the translation from SystemC 
to MASC as well as the programing restrictions that are required to ensure that the translation preserves semantics.  
For a comprehensive description of SystemC, the reader is referred the Language Reference Manual~\cite{systemc}.

\subsection{Numerical Data}

SystemC provides the following six class templates:

\begin{itemize}

\item {\tt sc\_uint<$n$>} and {\tt sc\_int<$n$>}: unsigned and signed integer types of width $n$, where $0 \leq n \leq 64$;

\item {\tt sc\_biguint<$n$>} and {\tt sc\_bigint<$n$>}: unsigned and signed integer types of arbitrary width $n \geq 0$;

\item {\tt sc\_ufixed<$n$,$m$>}and {\tt sc\_fixed<$n$,$m$>}: unsigned and signed fixed-point types of width $n$, with $m$ integer bits.

\end{itemize}
In the terminology of the reference manual, the {\tt sc\_uint<$n$>} and {\tt sc\_int<$n$>} types,
as well as the standard C numerical types, are {\it limited-precision}, and the other register types are {\it finite-precision}.
The latter are more general but less efficient than the former.
Finite-precision arithmetic in SystemC is performed in unbounded arbitrary-precision arithmetic and thus conforms to the MASC specification, 
whereas evaluation of expressions involving only limited-precision register types is performed in 64-bit modular arithmetic.

The correspondence between SystemC and MASC register types is clear, except that an integer type {\tt ui}$n$ (resp., {\tt si}$n$), 
for $n \leq 64$, may be implemented as either {\tt sc\_uint<$n$>} or {\tt sc\_biguint<$n$>} (resp., {\tt sc\_uint<$n$>} or 
{\tt sc\_biguint<$n$>}).  This choice should be made with the understanding that it is the SystemC programmer's responsibility to avoid 
arithmetic overflow and match the simpler semantics of MASC.

The MASC parser requires that all register variables be declared in MASC syntax.  Therefore, a program should begin by
declaring the register types that are used in the program, e.g.,

\begin{verbatim}
  typedef sc_biguint<81> ui81;
  typedef sc_fixed<16,4> sf16i4;
\end{verbatim}

Associated with the register class templates are (1)~an operator that has the same syntax and semantics as the MASC bit reference operator,
and (2)~a {\tt range} method of two arguments with C++ syntax but the same semantics as the MASC subrange operator.  The SystemC 
terms for expressions corresponding to a bit reference and a subrange are {\it bit-select} and {\it part-select}.

A straightforward syntactic conversion of part-selects to subrange expressions is performed by the MASC parser.  However, the
programmer should be aware of certain restrictions on the use of both bit-selects and part-selects.
For example, whereas a part-select of a limited-precision integer register
admits an implicit conversion to {\tt uint} and may therefore be treated as an object of that type, the return type of the {\tt range} method 
for finite-precision types admits no implicit conversion to any numerical 
type.  In order for such a value to be used in an arithmetic expression, it must first be explicitly converted to a 
{\tt uint} by the {\tt to\_uint} method, as in 
\begin{center}
{\tt x.range(5,3).to\_uint() + 2}.
\end{center}
Since the {\tt uint} value of this method is limited to 64 bits, it must not be applied to a part-select wider than 64 bits.

The SystemC rules regarding assignment between register types are complicated and will not be addressed here. (See~\cite{systemc}.)

\subsection{Arrays}

The stipulation that MASC arrays are passed only by value dictates that C arrays may be used only as global constants and 
in instances where arrays are used only locally by a function.  The effect, however, of passing arrays as
value parameters may be achieved by means of a simple C++ class template:
\begin{verbatim}
template <uint m, class T>
class array {
  public:
    T elt[m];
};
\end{verbatim}
This template is defined in the header file {\tt "books/imul/masc.h"}, which must be included in any SystemC 
model that is intended for translation to MASC.
As an illustration of its use, suppose that we define a function as follows:

\begin{verbatim}
  array<8, int> Sum8(array<8, int> a, array<8, int> b) {
    for (uint i=0; i<8; i++) {
      a.elt[i] += b.elt[i];
    }
    return a;
  }
\end{verbatim}
If {\tt a} and {\tt b} are variables of type {\tt array<8, int>}, then the result of the assignment
\begin{verbatim}
  b = Sum8(a, b);
\end{verbatim}
which does not affect the value of {\tt a}, is that each entry of the array {\tt b.elt} is incremented by the corresponding 
entry of {\tt a.elt}.

Aside from the restriction on parameter passing, there is no semantic difference between ordinary C arrays and
instances of an {\tt array} template class.  The MASC parser, therefore, simply converts {\tt array} objects to C arrays:
\begin{verbatim}
  int[8] Sum8(int a[8], int b[8]) {
    for (uint i=0; i<8; i++) {
      a[i] += b[i];
    }
    return a;
  }
\end{verbatim}

\subsection{Multiple Values}\label{mv}

A second class template is provided to achieve the effect of multiple-valued functions:
\begin{verbatim}
template <class T1, class T2, class T3 = bool, class T4 = bool>
class mv {
    T1 first; T2 second; T3 third; T4 fourth;
  public:
    mv(T1 x, T2 y) {first = x; second = y;}
    mv(T1 x, T2 y, T3 z) {mv(x, y); third = z;}
    mv(T1 x, T2 y, T3 z, T4 w) {mv(x, y, z); fourth = w;}
    void assign(T1 &x, T2 &y) {x = first; y = second;}
    void assign(T1 &x, T2 &y, T3 &z) {assign(x, y); z = third;}
    void assign(T1 &x, T2 &y, T3 &z, T4 &w) {assign(x, y, z); w = fourth;}
};
\end{verbatim}
A single function call may result in the assignment of values to as many as four variables by declaring the return type of 
the function to be {\tt mv<$T_1$,$\ldots$,$T_k$>}, where $2 \leq k \leq 4$ and $T_1,\ldots,T_k$ are arbitrary types, and 
applying the {\tt assign} method.  For example, the following is the C++ version of the {\tt Divide} function of 
Subsection~\ref{functions}:
\begin{verbatim}
  mv<uint, uint> Divide(m uint, n uint) {
    assert(n != 0);
    uint quot = 0, rem = m;
    while (rem >= n) {
      quot++;
      rem -= n;
    }
    return mv<uint, uint>(quot, rem);
  }
\end{verbatim}
A call to this function has the following syntax:
\begin{verbatim}
  uint q, r;
  Divide(23, 5).assign(q, r);
\end{verbatim}
Note that the {\tt assign} method is the only context in which we allow the use of reference parameters.
Again, translation to MASC syntax is straightforward.

\subsection{Bounded Iteration}

A {\tt while} loop, or a {\tt for} loop that does not comply with the format described in Subsection~\ref{control},
may be effectively translated to MASC provided that the programmer is able to specify an upper bound on the number of
iterations performed by the loop.  This is conveyed to the parser by a comment immediately preceding the loop:
\begin{alltt}
  // MASC: {\it bound} iterations
\end{alltt}
where {\it bound} may be any integer expression free of any variables that are assigned within the loop.  For example,
if the {\tt while} loop of the {\tt Divide} function above is preceded by
\begin{verbatim}
  // MASC: m iterations
\end{verbatim}
then the following MASC code is generated:

\begin{verbatim}
  for (uint _i=0; _i<m && rem > n; _i++) {
   quot++;
   rem -= n;
  }
\end{verbatim}

\section{Mapping MASC to ACL2}\label{acl2}

Translation of a MASC model from SystemC to ACL2 is performed in two steps: (1) the MASC parser generates a representation of the
model as a set of S-expressions, and (2) an ACL2 program converts this representation to an ACL2 program.

\subsection{Primitive functions}

Most of the MASC primitives correspond naturally to built-in ACL2 functions.  The rest are
implemented by a set of functions defined in the RTL library book {\tt "lib/masc"}, which is included in
any book generated by the MASC-ACL2 translator.

Several of these functions pertain to bit manipulation:

\begin{itemize}

\item ({\tt bits} $x$ $i$ $j$) is the bit slice $x[i:j]$.

\item ({\tt bitn} $x$ $i$) is the single bit $x[i]$ = ({\tt bits} $x$ $i$ $i$).

\item ({\tt setbits} $x$ $w$ $i$ $j$ $y$) is the result of replacing the bit slice $i:j$ of the bit vector $x$ of width $w$ with $y$. 

\item ({\tt setbitn} $x$ $w$ $i$ $y$) is ({\tt setbits} $x$ $w$ $i$ $i$ $y$).

\item ({\tt cat} $x_1$ $w_1$ $\ldots$ $x_k$ $w_k$), for any $k \geq 2$, is the concatenation of the bit slices $x_i[w_i-1:0]$, $i=1,\ldots,k$.

\end{itemize}
The following are used in connection with register values (as in the example in Subsection~\ref{parser}):

\begin{itemize}

\item ({\tt fl} $x$) is the greatest integer not exceeding the rational number $x$.

\item ({\tt intval} $w$ $x$) is the signed integer represented by the bit vector $x$ of width $w$.

\end{itemize}
A number of functions are required for the translation of the boolean-valued operators of C, in which values are compared with 0,
to ACL2, in which values are compared with {\tt NIL}. For example,

\begin{itemize}

\item ({\tt log<} $x$ $y$) is 1 if $x<y$ and 0 if not.

\item ({\tt if1} $x$ $y$ $z$) is $z$ if $x = 0$ and $y$ if not.

\end{itemize}
MASC arrays (as well as structures) are represented in ACL2 as alists.  The pair corresponding to an array entry associates its index with
its value.  The following primitives are defined:

\begin{itemize}

\item ({\tt ag} $i$ $a$) is the value of the array $a$ at index $i$.

\item ({\tt as} $i$ $x$ $a$) is the result of setting the value of the array $a$ at index $i$ to $x$.

\end{itemize}

\subsection{The MASC Parser}\label{parser}

The S-expression generated by the parser for a MASC function has the form
\begin{center}
  ({\tt DEFUNC} {\it name} ($\mbox{\it arg}_1$ $\ldots$ $\mbox{\it arg}_k$) {\it body})
\end{center}
where {\it name} is the name of the function, $\mbox{\it arg}_1, \ldots, \mbox{\it arg}_k$ are its formal parameters,
and {\it body} is an S-expression derived from its body, which is assumed to be
a statement block.  The parser generates an S-expression for each statement as follows:

\begin{itemize}

\item Statement block:  ({\tt BLOCK} $\mbox{\it stmt}_1 \ldots \mbox{\it stmt}_k$).

\item Sinple assignment:  ({\tt ASSIGN} {\it var} {\it term}).

\item Multiple-value assignment:  ({\tt MV-ASSIGN} ($\mbox{\it var}_1 \ldots \mbox{\it var}_k$) {\it term}),
where {\it term} corresponds to a call to a multiple-valued function.

\item Variable or constant declaration:  ({\tt DECLARE} {\it var} {\it term}) or ({\tt ARRAY} {\it var} {\it term}), where {\it term}
is optional.

\item Conditional branch:  ({\tt IF} {\it term} {\it left} {\it right}), where {\it left} is a block and {\it right} is either
  a block or {\tt NIL}.

\item Return statement :  ({\tt RETURN} {\it term}).

\item For loop:  ({\tt FOR} ({\it init test update}) {\it body}), where {\it init} is a declaration or an assignment,
{\it test} is a term, {\it update} is an assignment, and {\it body} is a statement block.

\item Switch statement:  ({\tt switch} {\it test} ($\mbox{\it lab}_1$ . $\mbox{\it stmts}_1$) $\ldots$ ($\mbox{\it lab}_k$ . $\mbox{\it stmts}_k$)),
where $\mbox{\it lab}_i$ is either an integer or a list of integers and $\mbox{\it stmts}_i$ is a list of statements.

\item Assertion: ({\tt ASSERT} {\it fn term}), where {\it fn} is the name of the function in which the assertion
  occurs and {\it term} is a term that is expected to have a non-zero value.

\end{itemize}
Note that variable types are not explicitly preserved in the translation.  Instead, they are used by the parser to inform the translation of terms.
Consider, for example, the MASC statement block
\begin{verbatim}
{ sf8i2 x = -145;
  ui8 y = 100, z = 3;
  z = y[4:2] * x; }
\end{verbatim}
In the evaluation of the expression on the right side of the final assignment, the type of {\tt x} dictates that its value is interpreted as
a signed rational with 6 fractional bits, and according to the type {\tt y} and {\tt z}, their assigned values must be truncated to 8 
integer bits.  Thus, the SystemC program that generates the above MASC code also generates the following S-expression:
\begin{verbatim}
(BLOCK (DECLARE X (BITS (* -145 (EXPT 2 6)) 7 0))
       (LIST (DECLARE Y (BITS 100 7 0))
             (DECLARE Z (BITS 3 7 0)))
       (ASSIGN Z
               (BITS (FL (* (BITS Y 4 2)
                            (/ (INTVAL 8 X) (EXPT 2 6))))
                     7 0)))
\end{verbatim}

\subsection{Generating ACL2 Code}

The ACL2 program that operates on the output of the MASC parser resides in {\tt "books/imul/trans\-late.lisp"}.
(The parser itself has not been made publicly available.)
The overall strategy of this program is to convert the body of a function to a nest 
of {\tt LET}, {\tt LET*}, and {\tt MV-LET} terms.  For each statement in the body, 
the translator generates the following:

\begin{itemize}

\item {\it ins}: a list of the variables whose values (prior to execution of the statement) are read by the statement;

\item {\it outs}: a list of the variables (non-local to the statement) that are written by the statement;

\item {\it term}: an expression of which (a)~the unbound variables are {\it ins}, and (b)~the value is a multiple value 
consisting of the updated values of the variables of {\it outs}, or a single value if {\it outs} is a singleton.

\end{itemize}
Each statement except the last corresponds to a level of the nest in which the variables of {\it outs} are bound to the
value of {\it term}, except that as an optimization to improve readability, adjacent {\tt LET}s are combined into a single {\tt LET}
or {\tt LET*} whenever possible.  The {\it term} of the final statement of the body becomes the body of the nest.

As a trivial (and nonsensical) example, the SystemC that generates the MASC function
\begin{verbatim}
uint foo(uint x, uint y, uint z) {
  uint u = y + z, v = u * x;
  <x, y, z> = bar(u, v);
  y = x > y ? 2 * u : v;
  if (x >= 0) {
    u = 2*u;
  }
  else {
    v = 3 * u;
  }
  if (x < y) {
    return u;
  }
  else {
  return y + v;
  }  
}
\end{verbatim}
also generates the corresponding ACL2 function
\begin{verbatim}
(DEFUN FOO (X Y Z)
       (LET* ((U (+ Y Z)) (V (* U X)))
             (MV-LET (X Y Z) (BAR U V)
                     (LET ((Y (IF1 (LOG> X Y) (* 2 U) V)))
                          (MV-LET (V U)
                                  (IF1 (LOG>= X 0) 
                                       (MV V (* 2 U))
                                       (MV (* 3 U) U))
                                  (IF1 (LOG< X Y) U (+ Y V)))))))
\end{verbatim}

Assertions, which do not affect any program variables, are handled specially.  An assertion ({\tt ASSERT} {\it fn term}) 
results in a binding of the dummy variable {\tt ASSERT} to the value ({\tt IN-FUNCTION} {\tt fn term}), where 
{\tt IN-FUNCTION} is a macro, defined in {\tt "lib/masc.lisp"}, that throws an error if the value of {\it term} is 0, with a 
message indicating the function in which the error occurred.

In addition to the top-level ACL2 function corresponding to a MASC function, a separate 
recursive function is generated for each {\tt for} loop.  Its returned values are those of the non-local variables
that are assigned within the loop.  Its arguments include these variables, along 
with any variables that are required in the execution of the loop, as well as any variables 
that occur in the loop initialization or test.  The construction of this function is 
similar to that of the top-level function, but the final statement of the loop body is not 
treated specially.  Instead, the body of the nest of bindings is a recursive call in which 
the loop variable is replaced by its updated value.  The resulting term becomes the left 
branch of an {\tt IF} expression, of which the right branch is simply the returned variable 
(if there is only one) or a multiple value consisting of the returned variables (if there 
are more than one).  The test of the {\tt IF} is the test of the loop.

For example, the function
\begin{verbatim}
uint baz(uint x, uint y, uint z) {
  uint u = y + z, v = u * x;
  for (uint i=0; i<u && u < v; i+=2) {
    v--;
    for (int j=5; j>=-3; j--) {
      assert(v > 0);
      u = x + 3 * u;
    }
  }
  return u + v;
}
\end{verbatim}
generates three ACL2 functions:
\begin{verbatim}
(DEFUN BAZ-LOOP-0 (J V X U)
       (DECLARE (XARGS :MEASURE (NFIX (- J (1- -3)))))
       (IF (AND (INTEGERP J) (>= J -3))
           (LET ((ASSERT (IN-FUNCTION BAZ (> V 0)))
                 (U (+ X (* 3 U))))
                (BAZ-LOOP-0 (- J 1) V X U))
           U))

(DEFUN BAZ-LOOP-1 (I X V U)
       (DECLARE (XARGS :MEASURE (NFIX (- U I))))
       (IF (AND (INTEGERP I) (INTEGERP U) (INTEGERP V)
                (AND (< I U) (< U V)))
           (LET* ((V (- V 1)) (U (BAZ-LOOP-0 5 V X U)))
                 (BAZ-LOOP-1 (+ I 2) X V U))
           (MV V U)))

(DEFUN BAZ (X Y Z)
       (LET* ((U (+ Y Z)) (V (* U X)))
             (MV-LET (V U)
                     (BAZ-LOOP-1 0 X V U)
                     (+ U V))))
\end{verbatim}
\section{Case Study: An Integer Multiplier}\label{imul}

The multiplier model may be found in the ACL2 directory
{\tt "books/workshops/2014/russinoff-oleary/"}, which comprises the following files:

\begin{itemize}

\item {\tt README}: Instructions for generating, certifying, and testing the other files in the directory;

\item {\tt imul.cpp}:  A hand-coded SystemC representation of the multiplier, typical of the models written by architects of
arithmetic units;

\item {\tt imul.m}: The more readable and abstract MASC version of the model, generated by the MASC parser;

\item {\tt imul.ast.lsp}: The intermediate S-expression representation of the model, also generated by the MASC parser;

\item {\tt translate.lisp}: The ACL2 component of the MASC-ACL2 translator;

\item {\tt imul.lisp}: The ACL2 model generated from {\tt imul.ast.lsp} by {\tt translate.lisp};

\item {\tt proof.lisp}:  The script of a proof of correctness of the model.

\end{itemize}
This section presupposes access to this directory as well as the RTL library {\tt "books/rtl/rel9/lib/"}.

\subsection{Underlying Theory}

Booth encoding is a technique for reducing the number of partial products that must be summed in computing the product of two bit vectors.
While the naive approach leads to as many partial products as the bit-width of the multiplier, radix-4 Booth encoding reduces this
number by half. 

The details of the analysis outlined below may be found in the subsection of the library reference
manual~\cite{libman} entitled ``Radix-4 Booth Encoding'', which also relates this presentation to the ACL2 formalization
found in the library book {\tt "lib/mult"}: the functions $\theta$, {\it bmux4}, {\it pp4}, $S$, $\mbox{\it pp4}'$, and $S'$ 
defined below are formalized by the ACL2 functions {\tt theta}, {\tt bmux4}, {\tt pp4-theta}, {\tt sum-pp4-theta}, {\tt pp4p-theta}, and 
{\tt sum-pp4p-theta}, and the main result is the lemma {\tt booth4-corollary-2}.

As a notational convenience, we shall assume that $x$ and $y$ are bit vectors of widths $n-1$ and $2m-1$, respectively.  Our
objective is an efficient computation of $xy$ as a sum of $m$ partial products.  Conceptually, the multiplier $y$ is partitioned
into $m$ 2-bit slices, $y[2i+1:2i]$, $i = 0,\ldots,m-1$.  Corresponding to each slice, we define an integer $\theta_i$ in the
range $-2 \leq \theta_i \leq 2$, as
\[
\theta_i = theta_i(y) = y[2i-1] + y[2i] - 2y[2i+1].
\]
A simple inductive proof shows that
\[
xy = x\sum_{i=0}^{m-1}2^{2i}\theta_i = \sum_{i=0}^{m-1}2^{2i}x\theta_i.
\]
Each term of this sum will correspond to a partial product, constructed by means of a 5:1 multiplexer:
\[
\mbox{\it bmux4}_i = \mbox{\it bmux4}_i(x, y, n) = \left\{\begin{array}{ll}
     x             & \mbox{if $\theta_i = 1$}\\
     \verb!~!x[n-1:0]    & \mbox{if $\theta_i = -1$}\\
     2x            & \mbox{if $\theta_i = 2$}\\
     \verb!~!(2x)[n-1:0] & \mbox{if $\theta_i = -2$}\\
     0             & \mbox{if $\theta_i = 0$.}\end{array}\right.
\]
Thus, if $\theta_i \neq 0$, $\mbox{\it bmux4}_i$ is the $n$-bit value computed by (1)~shifting $x$ one bit left if $|\theta_i| = 2$
and (2)~taking the bit-wise complement if $\theta_i < 0$.  (Inspection of the corresponding ACL2 definition reveals that the complement
operator, which is defined simply as \verb!~!$x = 1-x$, takes preference over the bit extraction operator, so that, for example,
since $x$ is an $(n-1)$-bit vector, \verb!~!$x[n-1:0]$ is an $n$-bit vector with leading bit 1.)

Now for $i = 0,\ldots,m-1$, let $B_i = B_i(x, y, n) = \mbox{\it bmux4}(\theta_i, x, n)$ and
\[
\mbox{\it neg}_i = \mbox{\it neg}_i(y) = \left\{\begin{array}{ll}
  0 & \mbox{if $\theta_i \geq 0$}\\
  1 & \mbox{if $\theta_i < 0$.}\end{array}\right.
\]
If we define the $(n+2m)$-bit aligned partial products (using a Verilog-inspired notation for concatenation) by
\[
\mbox{\it pp4}_i = \mbox{\it pp4}_i(x, y, m, n) = \left\{\begin{array}{ll}
  \{2(m\!-\!1)\verb!'b0!, \verb!1'b1!, \verb!~!\mbox{\it neg}_0, B_0[n\!-\!1:0]\} & \mbox{if $i=0$}\\
  \{2(m\!-\!i\!-\!1)\verb!'b0!, \verb!1'b1!, \verb!~!\mbox{\it neg}_i, B_i[n\!-\!1:0], 0'b0, \mbox{\it neg}_{i-1}, 2(i\!-\!1)\verb!'b0!\} & \mbox{if $i \neq 0$,}
\end{array}\right.
\]
and form the sum $S = S(x, y, m, n) = \sum_{i=0}^{m-1}\mbox{\it pp4}_i$, then it is not difficult to show that
\[
2^n + S = 2^{n+2m} + xy.
\]
On the other hand, if we modify the definition of $\mbox{\it pp4}_0$ as follows,
\[
\mbox{\it pp4}'_i = \left\{\begin{array}{ll}
  \{2(m-2)\verb!'b0!, \verb!1'b1!, \verb!~!\mbox{\it neg}_i, \mbox{\it neg}_i, \mbox{\it neg}_i, B_0[n-1:0]\} & \mbox{if $i=0$}\\
  \mbox{\it pp4}_i & \mbox{if $i \neq 0$,}\end{array}\right.
\]
and let $S' = \sum_{i=0}^{m-1}\mbox{\it pp4}'_i$, then $\mbox{\it pp4}'_0 = \mbox{\it pp4}_0 + 2^n$ and we have
\[
S'[n+2m-1:0] = xy.
\]  
The ACL2 formalization of this result is
\begin{verbatim}
(defthm booth4-corollary-2
    (implies (and (not (zp n))
                  (not (zp m))
                  (bvecp x (1- n))
                  (bvecp y (1- (* 2 m))))
             (= (bits (sum-pp4p-theta x y m n) (1- (+ n (* 2 m))) 0)
                (* x y)))
  :rule-classes ())
\end{verbatim}
Our MASC multiplier is essentially an implementation of the function {\tt sum-pp4p-theta}, and the lemma {\tt booth4-corollary-2}
will be the basis of our proof of correctness.  In this application, both operands are of width 32, and therefore we have 
$n = 33$ and $m = 17$.

\subsection{Implementation and Proof Summary}

In this discussion, we shall reproduce the MASC code in the file {\tt "imul.m"} and refer to the corresponding ACL2 code
in {\tt "imul.lisp"} and the lemmas in {\tt "proof.lisp"}.  References to mechanically generated ACL2 code are in upper case, and those to
hand-written code are in lower case.  We adhere to the notation established in the preceding subsection.

The first step of the computation is the construction of an array of encodings of the 17 Booth digits $\theta_i$,
involving two MASC functions:

\begin{verbatim}
ui3 Encode(ui3 slice) {
  ui3 enc;
  switch (slice) {
  case 4: enc = 6; break;
  case 5: case 6: enc = 5; break;
  case 7: case 0: enc = 0; break;
  case 1: case 2: enc = 1; break;
  case 3: enc = 2; break;
  default: assert(false);
  }
  return enc;
}

ui3[17] Booth(ui32 x) {
  ui35 x35 = x << 1;
  ui3 a[17];
  for (int k = 0; k < 17; k++) {
    a[k] = Encode(x35[2 * k + 2:2 * k]);
  }
  return a;
}
\end{verbatim}
These correspond to the three ACL2 functions {\tt ENCODE}, {\tt BOOTH-LOOP-0}, and {\tt BOOTH}.  The first of these
is characterized by the lemma {\tt encode-lemma}, proved by GL, which states that if its argument is the bit slice $2k+2:2k$ of the
shifted multiplier $2x$, then the value returned is a three-bit representation of the corresponding Booth digit $\theta_i$,
consisting of a sign bit and a two-bit magnitude.

The function {\tt BOOTH-LOOP-0} provides a simple illustration of one of the main problems faced in ACL2 verification of
recursive functions derived from {\tt for} loops: two inductions are required by the lemmas {\tt booth-recursion-1} and 
{\tt booth-recursion-2}, the second of which gives the desired characterization of the function.  The derived result 
{\tt booth-lemma} is a simple instantiation of {\tt booth-recursion-2}, stating that the value of {\tt BOOTH} is an array
of the values generated by {\tt ENCODE}.

The next step is the computation of $\mbox{\it bmux4}_i$:

\begin{verbatim}
ui33[17] PartialProducts(ui3 m21[17], ui32 x) {
  ui33 pp[17];
  for (int k = 0; k < 17; k++) {
    ui33 row;
    switch (m21[k][1:0]) {
    case 2: row = x << 1; break;
    case 1: row = x; break;
    default: row = 0;
    }
    pp[k] = m21[k][2] ? ~row : row;
  }
  return pp;
}
\end{verbatim}
Analysis of this function again involves two inductions, leading to the result {\tt partialprod\-ucts-lemma},
proved by GL, stating that the $k^{\text {th}}$ entry of the array returned by {\tt PARTIALPRODUCTS} is
$\mbox{\it bmux4}_k$.

Next, the aligned partial products are computed:

\begin{verbatim}
ui64[17] Align(ui3 bds[17], ui33 pps[17]) {
  bool sb[17], psb[18];
  for (int k = 0; k < 17; k++) {
    sb[k] = bds[k][2];
    psb[k + 1] = bds[k][2];
  }
  ui64 tble[17];
  for (int k = 0; k < 17; k++) {
    ui67 tmp = 0;
    tmp[2 * k + 32:2 * k] = pps[k];
    if (k == 0) {
      tmp[33] = sb[k];
      tmp[34] = sb[k];
      tmp[35] = !sb[k];
    }
    else  {
      tmp[2 * k - 2] = psb[k];
      tmp[2 * k + 33] = !sb[k];
      tmp[2 * k + 34] = 1;
    }
    tble[k] = tmp[63:0];
  }
  return tble;
}
\end{verbatim}
The ACL2 translation of this function includes a recursive function corresponding to each of two loops.
The first of these, according to {\tt sign-bits-lemma}, returns two arrays, the $k^{\text {th}}$ entries
of which are $\mbox{\it neg}_k$ and $\mbox{\it neg}_{k-1}$, respectively.  The second loop inserts these
bits into the aligned partial products.  The lemma {\tt align-lemma}, which is again proved by GL, 
states that the $k^{\text {th}}$ entry of the array computed by {\tt ALIGN} is the bit slice $63:0$ of 
$\mbox{\it pp4}'_k$.  This result is combined with the library lemma {\tt booth4-corollary-2} in the
proof of {\tt sum-simple-align-prod}.  This lemma refers to the function {\tt sum-simple}. which is a
straightforward computation of the
64-bit sum of an initial segment of an array of bit vectors.  It states that the 64-bit sum of the 17
entries of the array returned by {\tt ALIGN} is the product $xy$.  Note that in addition to {\tt booth4-corollary-2},
its proof requires hints pertaining to three lemmas from the book {\tt "lib/bits"}.

The final step is the computation of the sum.  This is implemented as a 17:2 compression tree followed 
by a single 64-bit addition:

\begin{verbatim}
ui64 Sum(ui64 in[17]) {
  ui64 A1[8];
  for (uint i = 0; i < 4; i++) {
    <A1[2*i+0], A1[2*i+1]> = Compress42(in[4*i], in[4*i+1], in[4*i+2], in[4*i+3]);
  }
  ui64 A2[4];
  for (uint i = 0; i < 2; i++) {
    <A2[2*i+0], A2[2*i+1]> = Compress42(A1[4*i], A1[4*i+1], A1[4*i+2], A1[4*i+3]);
  }
  ui64 A3[2];
  <A3[0], A3[1]> = Compress42(A2[0], A2[1], A2[2], A2[3]);
  ui64 A4[2];
  <A4[0], A4[1]> = Compress32(A3[0], A3[1], in[16]);
  return A4[0] + A4[1];
}
\end{verbatim}
The basic components of the tree are a 4:2 and a 3:2 compressor.  It consists of four levels, composed of 
four 4:2 compressors, two 4:2 compressors, one 4:2 compressor, and one 3:2 compressor, respectively.  The complexity 
of this design apparently exceeds the limits of GL, but the functionality of the components as specified by the 
GL lemmas {\tt compress42-lemma} and {\tt compress32-lemma} is readily verified.  

The function {\tt SUM} illustrates another common difficulty in verifying ACL2 functions derived from an
imperative language, which tend to be lengthy and lacking in modularity.  This can generally be effectively addressed
by establishing an equivalent modular formulation of the function.  In this case, this is accomplished by the lemma
{\tt sum-rewrite}, which effectively separates the stages of computation.  Through systematic application of
{\tt compress42-lemma} and {\tt compress32-lemma}, we derive results characterizing each level of the tree, culminating 
in the lemma {\tt sum-sum-simple}, which states the correctness of function {\tt SUM}.

The top-level function is a simple composition of the steps enumerated above:

\begin{verbatim}
ui64 Imul(ui32 s1, ui32 s2) {
  ui3 bd[17] = Booth(s1);
  ui33 pp[17] = PartialProducts(bd, s2);
  ui64 tble[17] = Align(bd, pp);
  ui64 prod = Sum(tble);
  return prod;
}
\end{verbatim}
Our final result follows immediately from {\tt sum-sum-simple} and {\tt sum-simple-align-prod}:

\begin{verbatim}
(defthm imul-thm
  (implies (and (bvecp s1 32) (bvecp s2 32))
           (equal (imul s1 s2) (* s1 s2))))
\end{verbatim}

\section{Conclusion}\label{conc}

At Intel we have applied MASC to several interesting and non-trivial examples.
Our most ambitious effort to date has been the development and modeling of the
division and square root algorithm mentioned in Section~\ref{intro}.
A MASC model of this design served as a 
platform for development and evaluation of new algorithmic features, optimizations, 
and modes of operation.  The same model has been deployed to RTL design and validation 
teams as an unambiguous, executable reference specification, and has
proved invaluable as an aid to understanding the algorithm and its realizations,
and in enabling RTL reuse and debugging.  The ACL2 version has been used extensively.
in testing variations of the algorithm against formally verified library functions,
exposing several bugs in the original design.  A comprehensive proof of correctness 
has been mechanically verified,

MASC has also found use as a tool for rapidly exploring and
evaluating new designs.  For example, a novel implementation of a new
combinatorial machine instruction was modeled in MASC by its architect,
and we were able to deliver an ACL2 proof of correctness in a few days.
We also exploited MASC's path to synthesizable SystemC to
obtain initial RTL code, enabling early evaluation of the area and
power characteristics of the design.

These applications have confirmed the expected benefits of combining the efficient
executability of C++, the expressiveness and synthesizability of SystemC, the simple
clarity of MASC, and the power of the ACL2 prover.  They have also been
instrumental in the development of an effective verification methodology.  The divider 
in particular played an important role in addressing the difficulties inherent in the 
analysis of ACL2 functions derived from inperative programs (which were touched upon 
in Section~\ref{imul} in connection with the simpler integer multiplier) as well as
the complications presented by fixed-point registers (which were not).  It also 
illustrates the modeling of a sequential circuit in MASC using iteration.  The 
combinatorial circuit mentioned above provides a compelling example of the power of 
our theorem proving approach in conjunction with the symbolic simulation capabilities 
of GL.  We plan to describe these applications in detail in future publications.

The MASC project has begun to achieve its motivating objectives
with respect to formal analysis, documentation, and RTL implementation,
and we have been gratified by the eagerness with which the language and methodology have
been adopted by architects and designers.  One interesting area for future work is the
potential use of MASC models and their correctness proofs in formal RTL verification. 

\nocite{*}
\bibliographystyle{eptcs}
\bibliography{ref}
\end{document}